\documentclass[12pt,preprint]{aastex}
\newcommand{\eg}{{\it e.g.}\ }

\newcommand{\etal}{{\it et al.}\ }

\def\gae{\lower 2pt \hbox{$\, \buildrel {\scriptstyle >}\over {\scriptstyle \sim}\,$}}
\def\lae{\lower 2pt \hbox{$\, \buildrel {\scriptstyle <}\over {\scriptstyle \sim}\,$}}
\begin{document}

%\title{Asteroid age distributions determined by colors and dynamics}

%\author{Mark Willman$^1$, Robert Jedicke$^1$, Bill Bottke$^2$, Nicholas Moskovitz$^1$}

\begin{center}

\vskip 8cm

{\bf Asteroid age distributions determined by space weathering\\
 and collisional evolution models\\
}

\vskip 2cm

{Mark Willman, Robert Jedicke}

\vskip 2cm

{Institute for Astronomy,  University of Hawai`i at Manoa\\
2680 Woodlawn Drive, Honolulu, HI{\ \ }96822}\\
{willman@ifa.hawaii.edu, 808-956-6989 tel, 808-956-9580 fax\\
jedicke@ifa.hawaii.edu, 808-956-9841 tel, 808-956-9580 fax}\\

%\vskip 0.5 cm

%{$^2$Department of Space Studies, Southwest Research Institute\\
%1050 Walnut Street, Suite 400, Boulder, CO{\ \ }80302}\\
%{bottke@boulder.swri.edu, 303-546-9670 tel, 303-546-9687 fax}

\end{center}

\vskip 2.5 cm

\noindent 23 pages

\vskip 0.5cm

\noindent 5 figures

\vskip 0.5 cm

\noindent 1 table

\clearpage

\begin{center}
{\bf ABSTRACT}
\end{center}
%\begin{abstract}

We provide evidence of consistency between the dynamical evolution of
main belt asteroids and their color evolution due to space weathering.
The dynamical age of an asteroid's surface
\citep{bib.bot05a,bib.nes05} is the time since its last catastrophic
disruption event which is a function of the object's diameter.  The
age of an S-complex asteroid's surface may also be determined from its
color using a space weathering model
\citep[\eg][]{bib.wil10,bib.jed04,bib.wil08,bib.mar06}.  We used a
sample of 95 S-complex asteroids from SMASS and obtained their
absolute magnitudes and $u,g,r,i,z$ filter magnitudes from SDSS.  The
absolute magnitudes yield a size-derived age distribution.  The $u,g,r,i,z$
filter magnitudes lead to the principal component color which yields a
color-derived age distribution by inverting our color-age
relationship, an enhanced version of the `dual $\tau$' space
weathering model of \citet{bib.wil10}.

We fit the size-age distribution to the enhanced dual $\tau$ model and
found characteristic weathering and gardening times of $\tau_w = 2050
\pm 80$ Myr and $\tau_g = 4400^{+700}_{-500}$ Myr respectively.  The
fit also suggests an initial principal component color of $-0.05 \pm
0.01$ for fresh asteroid surface with a maximum possible change of the
probable color due to weathering of $\Delta PC = 1.34 \pm 0.04$.  Our
predicted color of fresh asteroid surface matches the color of fresh
ordinary chondritic surface of $PC_1 = 0.17 \pm 0.39$.

%\end{abstract}

\vskip 1cm

%\begin{keywords}

%\clearpage

\section{Introduction}
\label{s.introduction}

During the twentieth century the number of known main belt (MB)
asteroids jumped from less than 500 to nearly 20,000.  The mushrooming
inventory led to the discovery of asteroid families \citep{bib.hir18}
and the discovery of the colors of the main spectral types
\citep{bib.cha75}.  This was followed late in the century by the
creation of an extensive database of colors by the Sloan Digital Sky
Survey (SDSS \citep{bib.aba09}), making possible high statistics color
population studies of asteroids.  However, an outstanding missing
element in the understanding of asteroid evolution was a timeline ---
at least until dynamical methods were developed for estimating family
age \citep[\eg][]{bib.mar95,bib.vok06a,bib.vok06b,bib.nes02}.  The
combination of family ages and colors based on remote observations led
to the \citet{bib.jed04} relationship between an asteroid surface's
color and its age.  We will define the 'color' of an asteroid as a
linear combination of filter magnitudes.  It is correlated with
spectral slope \citep{bib.wil08}.  An independent measure of an
asteroid's age depends on its size --- the surface of a large asteroid
remains intact longer than a small one because it suffers fewer
catastrophic disruptions.

The age of an asteroid family can be determined by several dynamical
methods, including family size frequency distribution (SFD) modeling,
global MB SFD modeling, modeling of family spreading via thermal
forces, and backward numerical integration of orbits
\citep{bib.nes05}.  A combination of these methods provides age
estimates for about 20 S- and C- complex families.  

Until recently, few families were known to be younger than ten Myr.
Prior to 2006 there were only two such S-complex families, Karin and
Iannini, but in that year \citet{bib.nes06a} and \citet{bib.nes06b}
identified four small genetic clusters of asteroids aged $<$ 1 Myr.
Two years later \citet{bib.pra09b} and \citet{bib.vok08} discovered
even younger dynamical pairs of asteroids that separated $<$ 500 kyr
ago.  

The ensemble of all family asteroid ages as determined by orbital
dynamical calculations spans four orders of magnitude and is the first
factor required to understand the rate of space weathering using the
age-color relationship.

The second factor was color.  A conundrum for three decades has been
the mismatch between the spectra of the most common meteorites
(ordinary chondrites) and their most likely source (inner main belt
S-complex asteroids).  The space weathering hypothesis was postulated
\citep{bib.cha73} as a means of reconciling the bright, relatively
blue spectra with deep absorption bands of ordinary chondrites and the
dark, red spectra with shallow absorption bands of S-complex
asteroids.  It proposes that the surface colors of asteroids of the
same mineralogy will change in a systematic way with airless exposure
to the space environment.  

The rate of color change on S-complex asteroids has been measured only
within the last decade, providing models of surface reddening rate and
color range \citep{bib.wil10,bib.wil08,bib.nes05,bib.jed04}.  Space
weathering may be due to a combination of mechanisms that could
include solar protons or heavier ions, electrons, ultraviolet
radiation, micrometeorites, and cosmic rays (see
e.g. \citet{bib.cha04} and references therein).  However here, as in
\citet{bib.wil10}, we are primarily concerned with the phenomenology
of space weathering rather than its cause.

Some recent space weathering models have assumed that the amount of
unweathered surface will decay exponentially over time
\citep{bib.wil10,bib.jed04}.  One would expect this result if the flow
of the weathering agent was constant, an approximation that is
probably valid over long periods of time.  \citet{bib.wil10} assume
that at the nanometer scale a part of the surface is either
unweathered or weathered and that the two states have distinct colors.
The physical motivation for this assumption is the metallic iron film
deposited on surfaces of nanophase silicate grains under bombardment
by pulsed lasers \citep{bib.sas01} or ion sputtering
\citep{bib.loe09}.  

Even as an asteroid's surface ages, weathered surface is transformed
back to an unweathered state by regolith gardening at an entirely
different rate due to micrometeorites, impact ejecta, seismic shaking,
electrostatic levitation, etc.  With both weathering and gardening in
mind \citet{bib.wil10} developed their 'dual $\tau$` model to describe
the changing colors of S-complex asteroids as a function of age, the
time since the family was created in a catastrophic collision that
generated a fresh surface on all family members.  The name `dual
$\tau$' captures the usage of independent characteristic times for
both space weathering $\tau_w$ and regolith gardening $\tau_g$.  Their
model extended the single $\tau$ model of \citet{bib.wil08} and
\citet{bib.jed04} by including young clusters in the analysis and by
explicitly including the physics of regolith gardening.  The dual
$\tau$ model yielded exponential characteristic times for weathering
and gardening of $\tau_w = 960 \pm 160$ Myr and $\tau_g = 2000 \pm
290$ Myr respectively.

The dual $\tau$ weathering time is consistent with four other results.
\citet{bib.pie00} used colors of craters dated by radiometry and
cosmic ray exposure ages to determine that space weathering on the
moon happens within 100-800 Myr.  This corresponds to a space
weathering time of 600-4800 Myr in the MB assuming the cause is
primarily solar in origin and the effect drops off as $1/r^2$.  

Similarly, \citet{bib.vev96} found that craters on (243)~Ida are bluer
than their surrounding background terrain.  The craters correspond to
freshly exposed and unweathered regolith while other parts of the
asteroid's surface indicates an age of about 1 Gy \citep{bib.gre96}.
The wide range in diameters (a proxy for crater age) of blueish crater
suggests that the space weathering time must be long.

In lab experiments \citet{bib.sas01} measured a weathering time of 100
Myr at 1 AU based on laser bombardment of olivine samples (equivalent
to 600 Myr in the MB) and this result was confirmed by
\citet{bib.bru06a}.

However, discrepant results include \citet{bib.loe09} who measured a
weathering time of only 0.005 Myr at 1 AU based on He ion bombardment
of olivine powder.  \citet{bib.bru06b} summarize ion irradiation
experiments finding a reddening time scale of order 1 Myr.
\citet{bib.ver09} propose a two-stage process with the first
accounting for most of the weathering and occurring in $<$ 1 Myr.

An independent surface age estimate can be determined from an
asteroid's size and its probability of catastrophic disruption
\eg\citet{bib.bot05a}.  Asteroid surfaces are completely reset during
catastrophic disruptions by impactors with diameters that are at least
a few per cent of the target's diameter while smaller impactors will
only have local surface effects.  Thus, the rate of catastrophic
disruption of asteroids depends on their size frequency distribution
(SFD) --- the larger the asteroid the longer it will survive
catastrophic disruption.  The SFD can be determined with observations
for large objects or through simulating MB collisional evolution that
is constrained by the observational results \eg\citet{bib.bot05a}.
The interval between catastrophic disruptions as a function of size
can only be determined from simulations.

In this work we use a sample of asteroids with known sizes and colors
to fit a color-age distribution to the independent size-age
distribution using an `enhanced dual $\tau$' model.  There is no a
priori reason the two age distributions must match for any combination
of parameters.  However, we will show that the size-age and color-age
methods are consistent and support both our space weathering model and
the collisional evolutionary models.  Our enhanced dual $\tau$ model
will correctly predict the color of fresh ordinary chondritic (OC)
material.

\section{Data sample}
\label{s.data}

Our goal is to compare independent age distributions determined from
the sizes and colors of a sample of asteroids.  To isolate the effect
of space weathering we use only S-complex asteroids thus reducing the
inherent mineralogical variation of the sample but still including
OC-like objects as shown by \eg \citet{bib.gaf93} and
\citet{bib.mor96}.  Thus, we selected a sample of 97 S-complex
asteroids from the Small Main-belt Asteroid Spectroscopic Survey
(SMASS) \citet{bib.bus02b} that also have $u,g,r,i,z$ filter
magnitudes and absolute magnitudes, $H$, from the Sloan Digital Sky
Survey (SDSS) \citep{bib.ive02}.  SMASS provided moderate resolution
spectra and definitively identified these asteroids as members of the
S-complex.  While this sample is smaller than the one in our previous
work \citep{bib.wil10} it has the advantage that each member has
rigorous type identification instead of simply relying on family
membership.

We used the SDSS $u,g,r,i,z$ filter magnitudes to assign a color to
each asteroid from which we determined its color-based age
(color-age).  Following \citet{bib.nes05} the principal component color
for an SDSS asteroid is
\begin{equation}
PC_1 = 0.396(u-g-1.43) + 0.553(g-r-0.44) + 0.567(g-i-0.55) +
0.465(g-z-0.58).
\label{eq.PC1}
\end{equation}
\noindent \citet{bib.wil08} showed that $PC_{1}$ is correlated with
spectral slope where increasing $PC_{1}$ corresponds to redder
asteroids.  Almost all the sample members have $ \, 0.15 < PC_1 < 0.82
\, $ with two outliers near $PC_1 = 1.70$.  We exclude the outliers
because of their extreme colors leaving 95 sample members.  Leaving
the 2 objects in the sample has little impact on the final result.

We independently determined the asteroid's ages from their diameters
as derived from their absolute magnitudes.  The diameter, $D$, for
each asteroid was calculated \citep{bib.bot05a} from its absolute
magnitude, $H$, and albedo, $p_v$, where we used an average albedo of
$0.215 \pm 0.041$ from another sample of 93 S-complex asteroids from
\citet{bib.bow08}:
\begin{equation}
\frac{D}{\rm{km}} = \frac{1329}{\sqrt{p_v}} 10^{-H/5} = 2863 \times 10^{-H/5}.
\label{eq.diameter}
\end{equation}

\section{Color-ages from the dual $\tau$ model}
\label{s.inversion}

\citet{bib.wil10} characterized the changing color of an S-complex
asteroid's surface as a function of time with their dual $\tau$ model
\begin{equation}
PC_1(t) = PC_1(0) + \Delta PC_1[1 - U(t,\tau_w,\tau_g)]
\label{eq.color}
\end{equation}
\noindent where
\begin{equation}
U(t,\tau_w,\tau_g) = \frac{e^{-(\frac{1}{\tau_g} + \frac{1}{\tau_w})t}
+ \frac{\tau_w}{\tau_g}} {1 + \frac{\tau_w}{\tau_g}}.
\label{eq.unweathered}
\end{equation}
\noindent The four parameters, $PC_1(0)$, $\Delta PC_1$, $\tau_w$, and
$\tau_g$ were fit to the asteroids' colors, $PC_1$, and ages, $t$,
using the least squares method.  They found $PC_{1}(0) = 0.37 \pm
0.01$, the color of unweathered surface, $\Delta PC_{1} = 0.33 \pm
0.06$, the maximum possible color change, and $\tau_w = 960 \pm
160$~Myr and $\tau_g = 2000 \pm 290$, the characteristic weathering
and gardening times respectively.

We can invert eq. \ref{eq.color} to determine an S-complex asteroid's
surface age from its color;
\begin{equation}
T_c \; = \; \frac{-1}{(\frac{1}{\tau_g} + \frac{1}{\tau_w})} \;
\rm{ln}\left[1 - \left(\frac{\mathit{PC_{1}(t) -
PC_{1}(\rm{0})}}{\Delta \mathit{PC_{1}}}\right) \left(1 +
\frac{\tau_{w}}{\tau_{g}}\right)\right].
\label{eq.swinversion}
\end{equation}
\noindent The logarithm in
eq. \ref{eq.swinversion} disallows negative arguments that result
from color values above the upper limit of
\begin{equation}
PC_{1,max} \; = \; PC_{1}(0) \; + \; \frac{\Delta \mathit{PC_{1}}}{1 +
\frac{\tau_w}{\tau_g}},
\label{eq.colormax}
\end{equation}
\noindent and restricts the invertible color range to
[$PC_{1}(0),PC_{1,max}$] with a concomitant restriction in the range
of predicted surface ages.  This is not a mathematical artifact; in
the simple analytic model gardening forestalls net weathering at the
equilibrium color of $PC_{1,max}$ and asteroids can not be younger
than freshly exposed surface.  This exposes a fundamental problem with
the dual $\tau$ model --- fitting a function involves finding the
central tendency of the data which leaves outlying points beyond the
function range.  In our sample of 95 objects about $\frac{1}{3}$ of
the colors exceed the allowed range of the model.

Thus, we developed an enhanced model described in the next section
that is based upon a probability density function (PDF) relating color
and age.  The model retains the advantage of incorporating the
demonstrated color-age relationship while avoiding the inversion
singularities.

\section{Color-ages from the enhanced dual $\tau$ model}
\label{s.coloragepdf}

The enhanced dual $\tau$ model is a PDF given by 
\begin{equation}
z(t,PC_1) = \frac{N}{\sqrt{2 \pi \sigma_{c}^2}} \; 
\exp\biggl[ {- \frac{(PC_1 - PC_1(t))^2}{2 \sigma_{c}^2}} \biggr]
\label{eq.pdf}
\end{equation}
\begin{figure}[h!]
\centerline{\includegraphics[width=2.7in,angle=90]{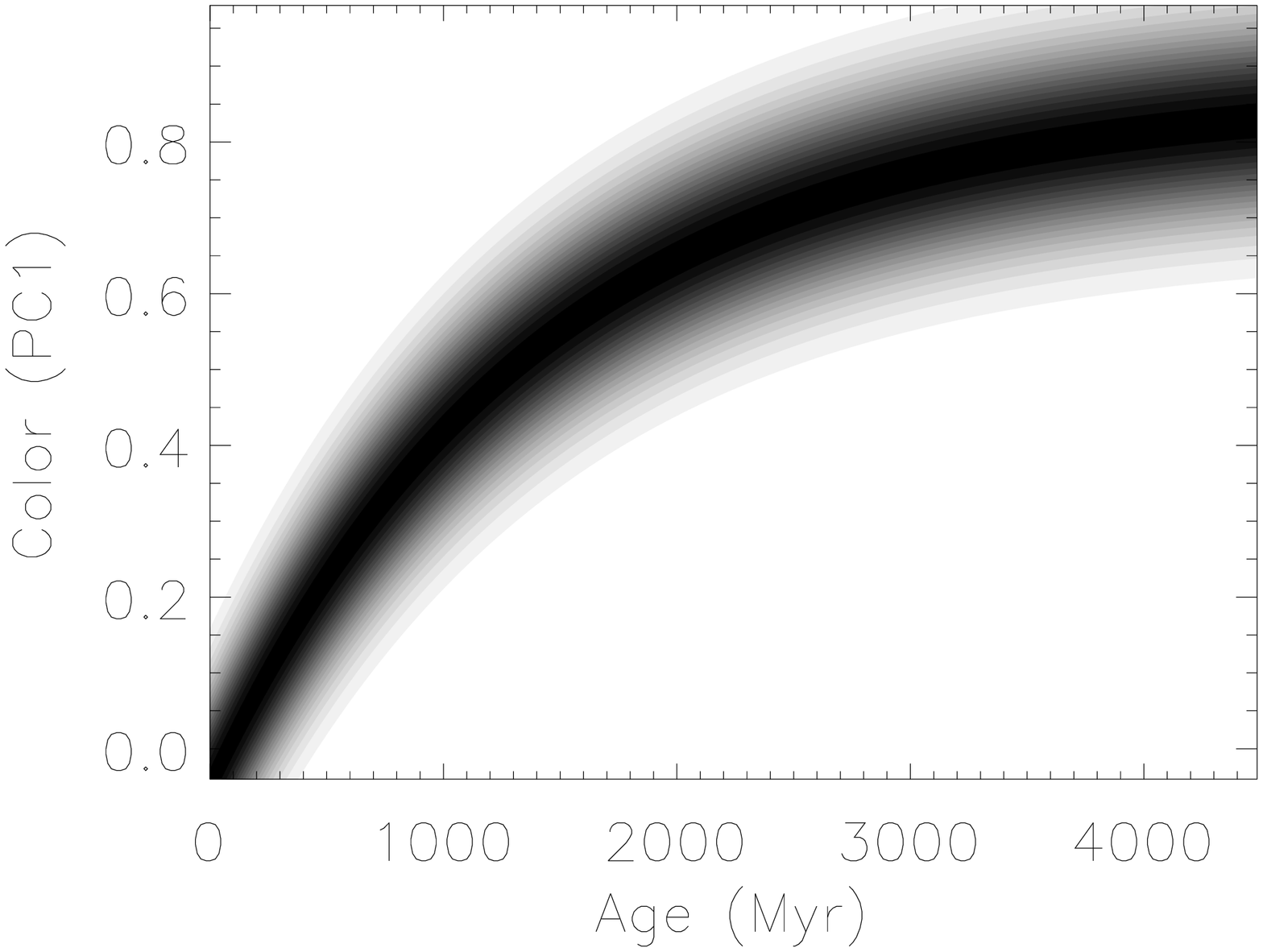}}
\centerline{\includegraphics[width=2.7in,angle=90]{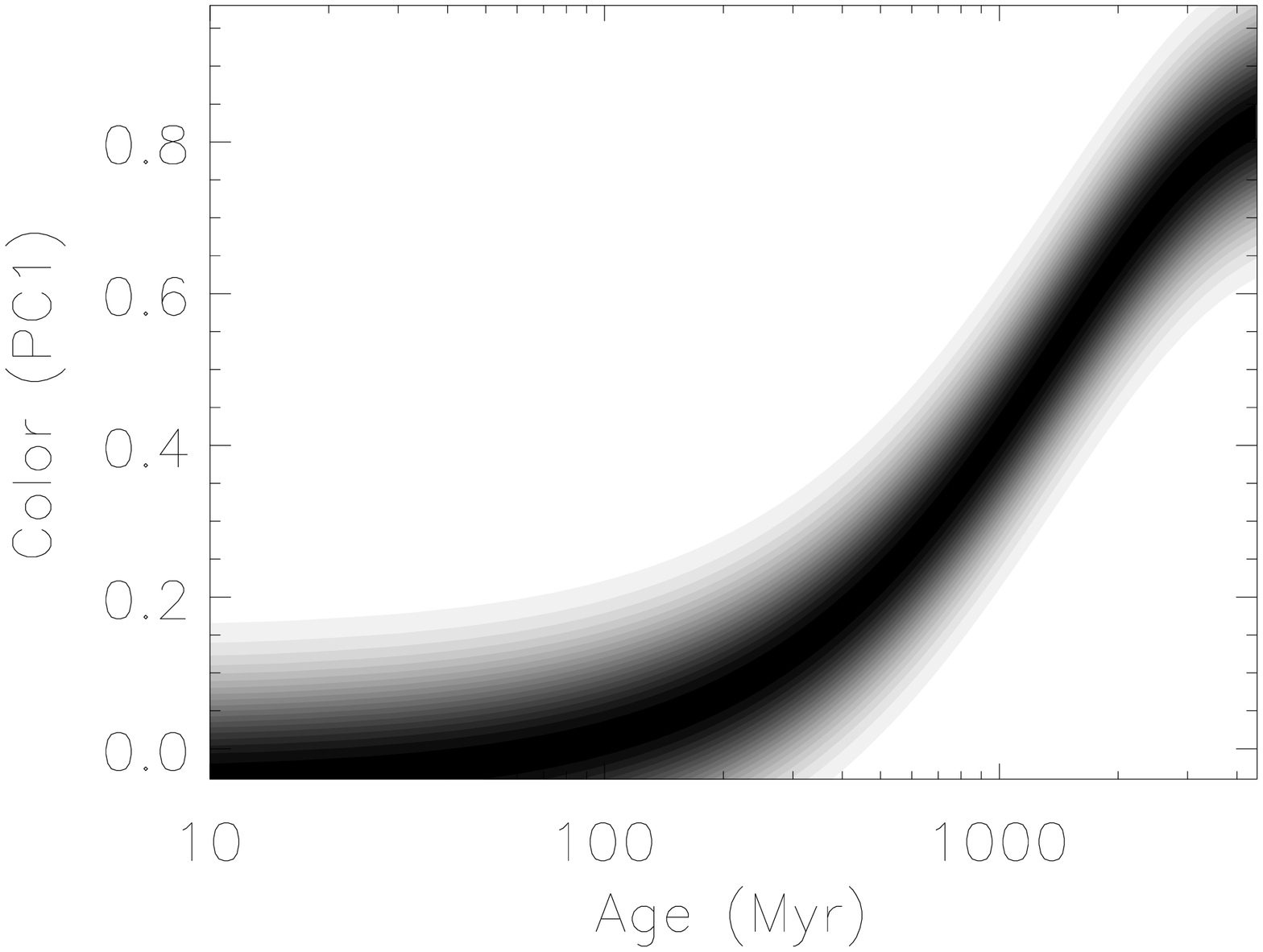}}
\caption{Top) The enhanced dual $\tau$ model of eq. \ref{eq.pdf} with
$\tau_w = 2050$ Myr, $\tau_g = 4400$ Myr, $PC_1(0)= -0.05$, and
$\Delta PC_1 = 1.34$.  This set of parameters is explained in
\S\ref{s.comparison}.  Darker regions correspond to higher
probability.  We use a linear time axis to facilitate the normal
interpretation of a pdf with equal areal density signifying equal
probability.  Bottom) The same function using a logarithmic time
scale.}
\label{f.pdf_contour_match}
\end{figure}
\noindent where $N$ is a normalization constant such that $\int\int z
\; dPC_1 \; dt = 1$, $PC_1(t)$ is the dual $\tau$ model of eqs
\ref{eq.color} and \ref{eq.unweathered}, and $PC_1$ and $t$ are free
parameters.  $PC_1 - PC_1(t)$ is the deviation from the predicted
color, $PC_1(t)$, for a given age.

To aid in interpreting the PDF the left side of Figure
\ref{f.pdf_contour_match} provides an example of the enhanced dual
$\tau$ model using a linear time scale --- each vertical cross-section
at a fixed age, $t$, is gaussian with its mean at $PC_1(t)$ and a
standard deviation in color of $\sigma_{c}$.

Our choice of $\sigma_{c}$ was motivated by the dual $\tau$ model that
was based on a fit to the average colors of twelve asteroid families.
The average rms spread of the families' colors had a mean of $0.085
\pm 0.017(rms)$ \citep{bib.wil10}.  Given the small deviation between
the asteroid families' rms spread we used a constant $\sigma_c =
0.085$.
\begin{figure}[h!]
\centerline{\includegraphics[width=5.1in,angle=90]{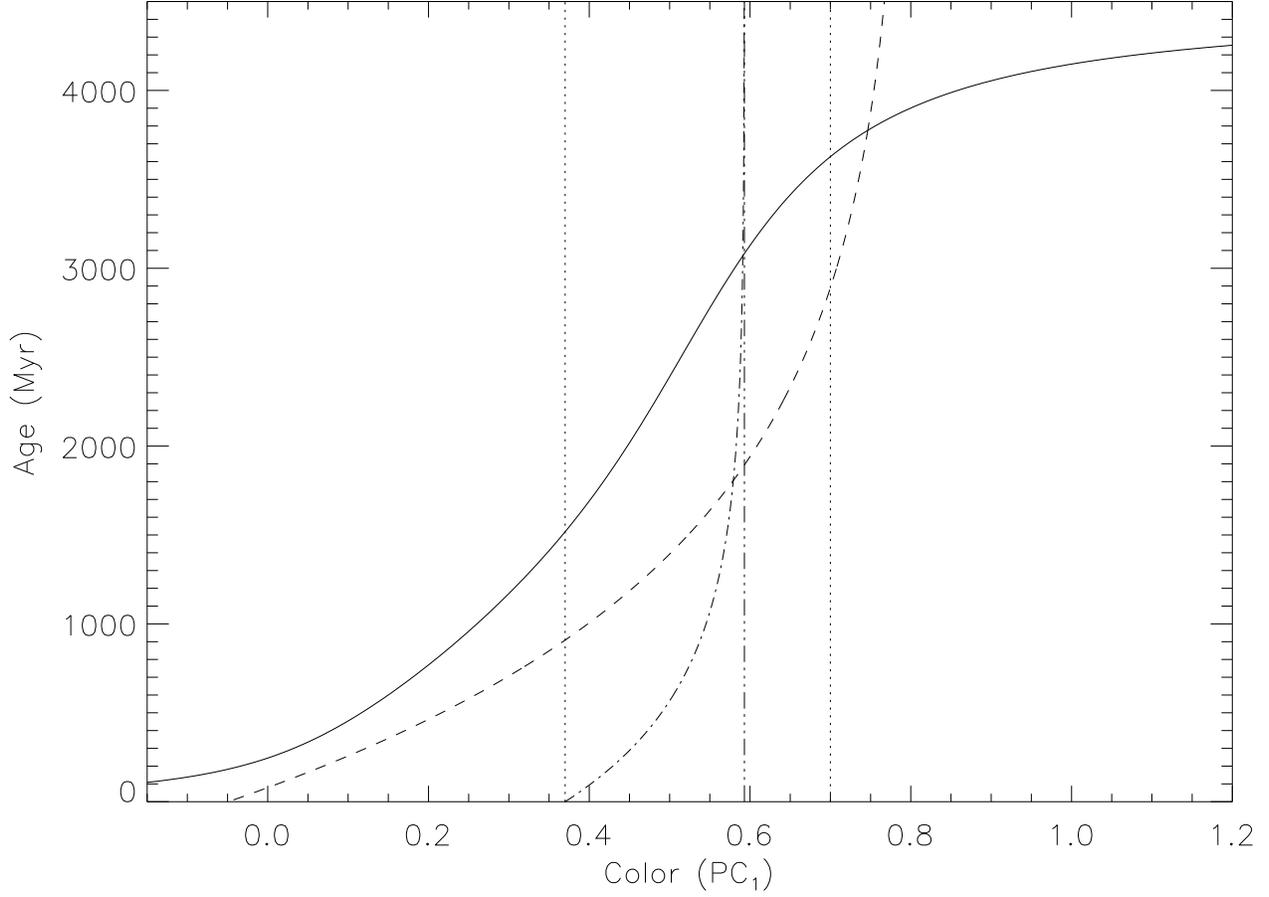}}
\caption{(dash dot) The color-age from the inverted dual $\tau$ model
  (eq. \ref{eq.swinversion}) which is undefined outside the dotted
  lines and further constrained at the (dash dot dot) asymptote by
  gardening. (dashed) The same function using enhanced dual $\tau$
  model parameters and (solid) the weighted mean color-age from
  eq.~\ref{eq.color_wt_age} that is defined for all colors.  The axes
  are flipped relative to Figure \ref{f.pdf_contour_match} because age
  is now considered a function of color instead of vice versa.}
\label{f.color_wt_age}
\end{figure}

Given an S-complex asteroid's $PC_1$ the best estimate for its age is
the weighted mean color-age
\begin{equation}
<T_c(PC_1)> \; = \frac{\int_0^{t_f} \; t \; z(t,PC_1) \; dt}{\int_0^{t_f} \; z(t,PC_1) \; dt}
\label{eq.color_wt_age}
\end{equation}
\noindent shown in Figure \ref{f.color_wt_age}.  Since $z(t,PC_1)$ is
defined for all colors as well as for ages dating back to the
beginning of the solar system, $t_f$, this color-age avoids the
inversion singularities of the dual $\tau$ model.

However, eq. \ref{eq.color_wt_age} does not account for the fact that
an asteroid $i$'s color includes a measurement error, $\Delta
PC_{1,i}$, that we model as a gaussian PDF:
\begin{equation}
s_i(PC_1) = \frac{1}{\sqrt{2 \pi \; (\delta PC_{1,i})^2}} \;
 e^{- \frac{(PC_1 - PC_{1,i})^2}{2 \; (\delta PC_{1,i})^2}}.
\label{eq.pdfi}
\end{equation}
\noindent Convolving this PDF with the enhanced dual $\tau$ PDF yields
the age dependent color-age PDF
\begin{equation}
T_{c,i}(t) = \frac{\int_{-\infty}^{\infty} \; z(t,PC_1) \; s_i(PC_1)
\; dPC_1} {\int_{0}^{t_f} dt \; \int_{-\infty}^{\infty} \; z(t,PC_1) \;
s_i(PC_1) \; dPC_1}.
\label{eq.colorage_pdf}
\end{equation}

Eq. \ref{eq.colorage_pdf} is not sensitive to $t_f$.  For instance,
the difference between using the age of the solar system (4.5 Gyr) and
the age of the oldest known asteroid family, Maria, at 3.0 Gyr,
results in a negligible shift of the PDF to slightly younger ages.
\begin{figure}[h]
\centerline{\includegraphics[width=5.0in,angle=90]{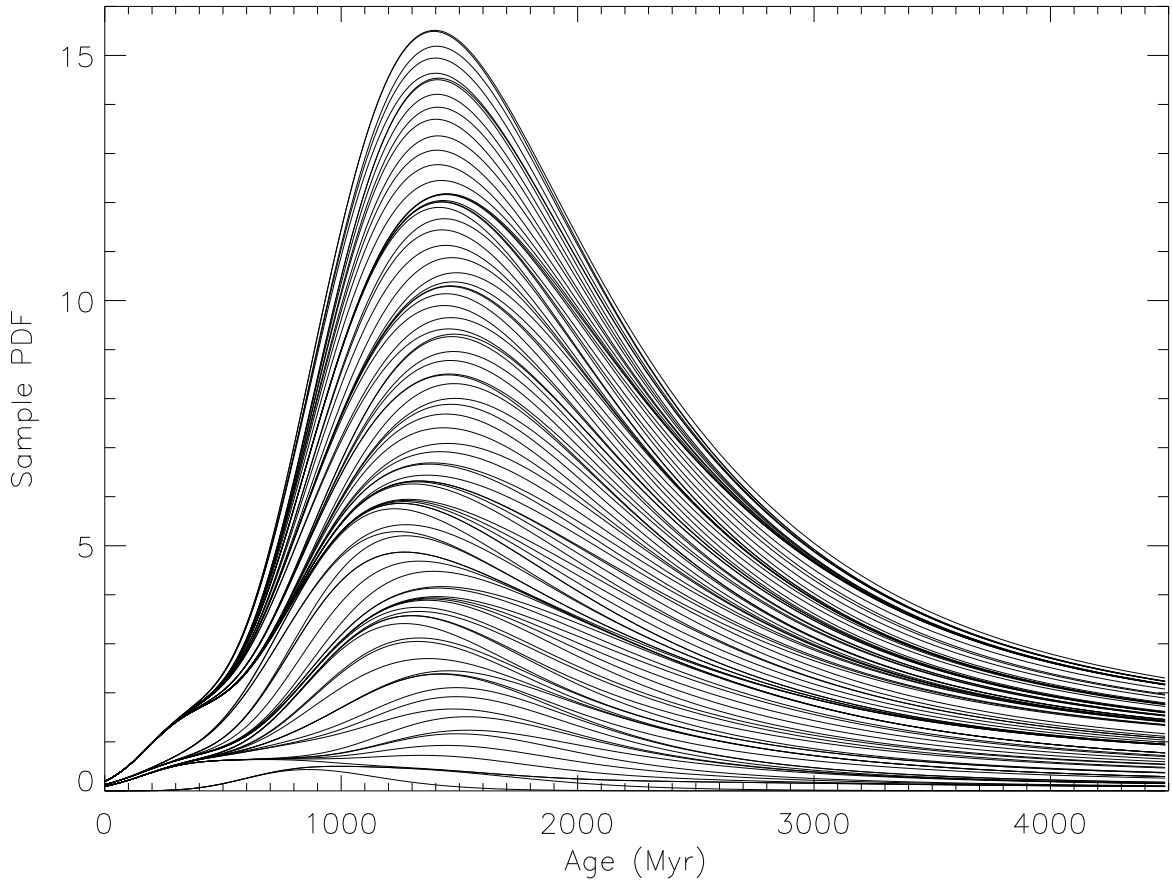}}
\caption{Superimposed (stacked) color-age PDFs from
eq. \ref{eq.colorage_pdf} for our sample of 95 S-complex asteroids.
We used the enhanced dual $\tau$ model parameters given in Table
\ref{t.params}.  The vertical scale is arbitrary.}
\label{f.superimposed_age_pdfs}
\end{figure}

Summing the color-age PDFs of all the asteroids in our sample yields
their combined differential color-age distribution
\begin{equation}
dN_c(t) \; = \; \displaystyle\sum_i \; T_{c,i}(t) \; dt ,
\label{eq.agedistn}
\end{equation}
\noindent the upper envelope of the curves shown in Figure
\ref{f.superimposed_age_pdfs}.  The envelope's maximum is near
1400~Myr but its mean lies near 2050~Myr due to the tail towards older
ages.  The envelope's mean is the analog of the weighted mean given in
eq. \ref{eq.color_wt_age} that corresponds to the enhanced dual $\tau$
model parameter $\tau_w$ in Table \ref{t.params}.

\section{Age distribution from asteroid sizes}
\label{s.sizeage}

\begin{figure}[h]
\centerline{\includegraphics[height=6.0in,width=4.5in,angle=90]{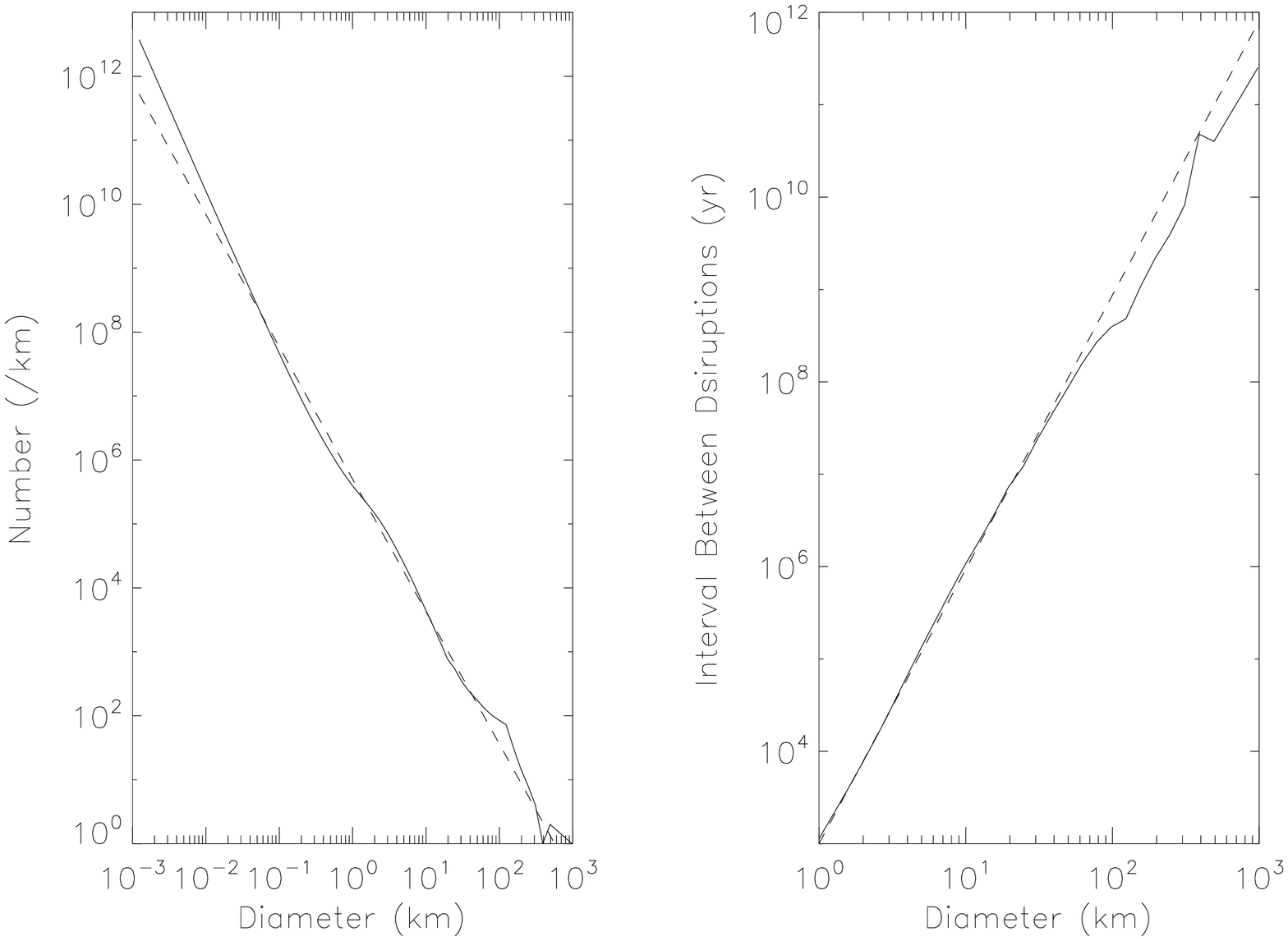}}
\caption{From \citet{bib.bot05a}.  Left: (solid) The MB differential
  SFD and (dashed) a power law fit. Right: (solid) The disruption
  interval as a function of diameter along with (dashed) a power law
  fit.}
\label{f.mbsfd}
\end{figure}
In the previous two sections we developed the enhanced dual $\tau$
model that enabled us to derive an age distribution from a sample of
asteroid colors.  We now develop an independent age distribution based
on asteroid sizes utilizing the fact that large asteroids are
resistant to catasrophic disruptions and therefore have older ages
than small asteroids.  While large asteroids may be rubble piles
reaccumulated in the aftermath of previous collisions
\citep[\eg][]{bib.mar95}, the age calculated here is the time since
the most recent catastrophic collision that last reset the asteroid's
surface age.

A common technique in simulating the MB's collisional evolution is to
numerically model the asteroids's collisional cascade.  The
simulations begin with an assumed initial SFD with asteroids
distributed across different diameter bins that then evolve as a
function of time accounting for the asteroids' specific energy as a
function of diameter.  An asteroid that suffers a collision disappears
from its bin and its fragments appear in their respective size bins.
The simulations are constrained by \eg the current MB SFD, Vesta's
intact surface, the number of large asteroid families, etc, and the
fidelity of the models encourages their use in determining the average
lifetime of asteroids as a function of diameter.

The average age, $\overline{T}$, of asteroids in a diameter bin is
half the average collisional lifetime in the bin, $\tau$, which in
turn is the mean time before destruction by catastrophic collision.
The average collisional lifetime is the product of the occupancy, $N$,
and disruption interval, $t_{dis}$, in the bin:
\begin{equation}
\overline{T} = \frac{\tau}{2} = \frac{N \; t_{dis}}{2}.
\label{eq.dynage}
\end{equation}
\noindent We obtained the quantities from the results of a simulation
by \citet{bib.bot05a} as shown in Figure \ref{f.mbsfd}.  

We used power law fits to $N$ and $t_{dis}$ to calculate the average
age as a function of diameter.  We restricted the fit to the diameter
range of 1-46 kilometers relevant to our 95 asteroids and find that $N
\sim 5 \times 10^5 \, (D/km)^{-2.07} \, dD/km$ where $dD$ is the
diameter bin width, and $t_{dis} \sim 1000 \, (D/km)^{2.97}$.
Combining eq. \ref{eq.diameter} with the last three equations results
in a size-derived age
\begin{equation}
T_s(\rm{yr}) = 3.23 \times 10^{(11.0 \, - \, 0.18 \, \mathit{H})}
\label{eq.agemag}
\end{equation}
\noindent where we use the observed absolute magnitude as a proxy for
diameter.

The size-age distribution for our sample using eq. \ref{eq.agemag} is
shown in Figure \ref{f.age_distns}.

\section{Results and discussion}
\label{s.comparison}

\begin{figure}[h]
\centerline{\includegraphics[width=4.7in,angle=90]{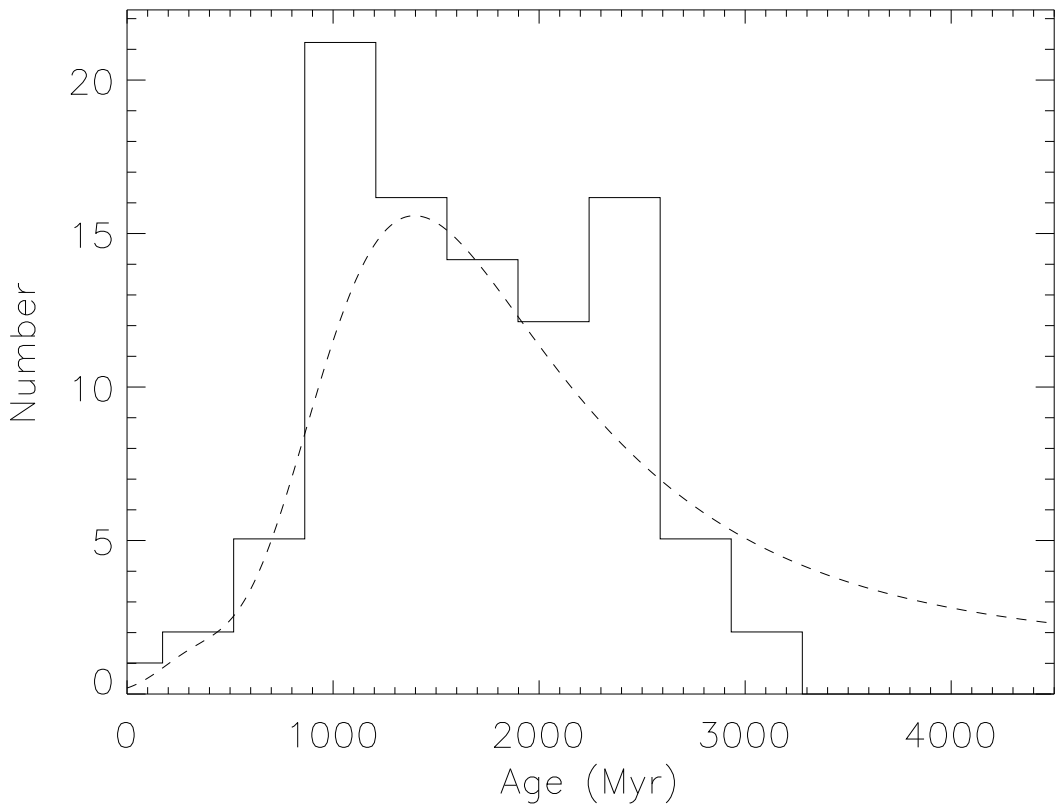}}
\caption{(solid) The size-age distribution from eq. \ref{eq.agemag}
  for our sample of 95 S-complex asteroids that are in both SDSS and
  SMASS, (dotted) the fit binned color-age distribution that yields
  the enhanced dual $\tau$ model, and (dashed) the resulting
  differential color-age distribution from eq. \ref{eq.agedistn}.}
\label{f.age_distns}
\end{figure}

In the previous two sections we described how to obtain an age
distribution based on asteroid color (color-age) as well as a diameter
dependent age distribution (size-age) derived from collisional
evolution models.  In this section we fit the color-age PDF to the
size-age distribution.  The fit yields the enhanced dual $\tau$
parameters given in Table \ref{t.params} and the function shown in
Figure \ref{f.age_distns}.

\begin{table}[h!]
\begin{center}
\begin{tabular}{ccccc}
\tableline\tableline
Model & $PC_1(0)$ & $\Delta PC_1$ & $\tau_w$ & $\tau_g$ \\
\tableline
Single $\tau$ & $0.31\pm0.04$ & $0.31\pm0.07$ & $570\pm220$ & na \\
Dual $\tau$ & $0.37\pm0.01$ & $0.33\pm0.06$ & $960\pm160$ & $2000\pm290$ \\
Enhanced dual $\tau$ & $-0.05 \pm 0.01$ & $1.34 \pm 0.04$ & $2050 \pm 80$ & $4400^{+700}_{-500}$ \\
\tableline\tableline
\end{tabular}
\end{center}
\caption{The space weathering model parameters for S-complex asteroids
derived from average family colors (single~$\tau$, \citet{bib.wil08};
dual~$\tau$, \citet{bib.wil10}) or from fitting the color-age PDF to
the size-age distribution (enhanced dual $\tau$).}
\label{t.params}
\end{table}

Our enhanced dual $\tau$ model is substantially different in
functional form, data sample, and derived parameters from the earlier
single and dual $\tau$ models.  The difference in functional form is
particularly relevant to the value and interpretation of the color
parameters, $PC_1(0)$ and $\Delta PC_1$.  The broadened range of the
color parameters reflect the fact that the PDF of the enhanced dual
$\tau$ model is not constrained in color and its inversion allows ages
to be assigned to asteroids of any color.  

We expect $PC_1(0)$, the most probable value for the initial color of
unweathered S-complex asteroids, to be bluer than the majority of
objects in the complex.  It should also match the color of freshly
exposed surfaces of ordinary chondrite meteorites in the laboratory.
The average color of 329 ordinary chondrite spectra\footnote{Of the
418 ordinary chondrites listed in \citet{bib.clo94} 329 included
spectra, had sufficient wavelength coverage, and were not irradiated.
We calculated a spectral slope of $0.10 \pm 0.46$/$\mu$m over the
SMASS standard wavelength range of $0.44-0.92 \mu$m and converted it
to $PC_1$ using the transformation derived in \citet{bib.wil08}.
\citet{bib.ver09} calculated a slope of 0.01/$\mu$m over a narrower
wavelength range that introduces a systematic offset of
$\sim$0.07/$\mu$m between their slope and this work.  Correcting their
value with this offset leads to only 0.02/$\mu$m difference in our
slope measurements.} from \citet{bib.clo94} is $PC_1 = 0.17 \pm
0.39(rms)$ in agreement with our new value of $PC_1(0)= -0.05 \pm
0.01$.  The large RMS on the mean meteorite color may be due to
differences in the meteorite's particle sizes and mineralogical
subtypes.  Thus it is difficult to make a meaningful comparison
between the meteorite colors and remotely sensed colors of asteroids
with unknown surface regolith structure.

Our long space weathering time of about 2000 Myr is surprisingly
different than \citet{bib.ver09}'s result of less than 1 Myr.  We are
unsure how to reconcile this three order of magnitude difference.
Their result relies on several corrections to the colors of asteroid
families and it is unclear what age was assigned to freshly cut
meteorite surfaces.  On the other hand, we fit our model to a sample
of less than 100 objects where most span the size range of about
$1-20$ km and an age range of about $100-3000$ Myr yet we predict, to
within 0.5 $\sigma$, the colors of meteorites that are orders of
magnitude removed in both size and age.  However we recognize that the
meteorites' RMS color variation is large.  

It is also possible that there are two weathering processes having
different time scales.  However, our data sample is not large enough
to justify exploring whether different weathering time scales are
present.  The decade of size range we use would not be sensitive to
the short timescales because the objects are too large to have
experienced recent catastrophic disruption.  Clearly, the resolution
of the discrepancy between the long and short weathering times
requires further work and insight.

In the absence of gardening $\Delta PC_1$ is the maximum possible
change of the most probable color of S-complex asteroid surfaces.
Gardening constrains the most probable maximum color from
eq. \ref{eq.colormax} to a lower value of $PC_1(0) + \Delta PC_{1}/(1
+ \frac{\tau_w}{\tau_g}) = 0.91 \pm 0.07$ as illustrated in Figure
\ref{f.color_wt_age}.  Figure \ref{f.pdf_contour_match} shows that
S-complex asteroids will not reach maximum redness even over the age
of the solar system --- consistent with our sample having a maximum
$PC_1 = 0.82$.

The dual and enhanced dual $\tau$ models give weathering times
differing by $\sim 6 \sigma$ as shown in Table \ref{t.params}.  However,
they were derived using independent data sets and different fit
techniques.  The dual $\tau$ model was fit to color vs. age data while
in this work we fit the inverted enhanced dual $\tau$ model to an age
distribution determined from sizes.  We believe the methodology of the
probabilistic enhanced dual $\tau$ model is superior because it avoids
the color truncation problem of the dual $\tau$ model, and attribute
the large difference in $\tau_w$ between the models to different data
sets and small sample size.  We expect that in the future larger data
samples with good colors along with better collisional evolution
models should resolve the discrepancy.  The main goal here was to
compare the consistency of ages from collisional evolution models and
the color inversion method.  The agreement of the age distributions in
Figure \ref{f.age_distns} accomplishes our goal.

\citet{bib.wil10} showed that the gardening time calculated there
(Table \ref{t.params}, dual $\tau$) was $\sim 7\times$ longer than the
resurfacing time calculated from impact rates and cratering physics.
The enhanced dual $\tau$ model gardening time is yet another $2\times$
longer!  The difference in the values is probably due to the same
reasons suggested above to account for the difference in weathering
times --- small sample size and the use of early collisional evolution
models to estimate the size-age of the asteroids.

The fidelity of our results to reality depends on the assumptions used
in this work.  For instance, we have assumed that $\sigma_c$, the
color width of the PDF as a function of time, is constant and equal to
a specific value determined from the RMS color distribution within
families, when it may actually vary with age.  With a larger sample
size and more accurate colors it may be possible to fit $\sigma_c$ as
a function of time.

Probably the largest systematic uncertainty in this work is the
assumption that the collisional evolution model is correct as we have
fit the color-age to the size-age distribution.  Those models depend
on several ill-determined inputs including but not limited to 1) the
current MB SFD 2) estimates of the initial SFD after accretion 3)
disruption, cratering, and ejecta physics 4) that material strength as
embodied in the specific disruption energy, $Q^*(D)$, is independent
of mineralogy 5) and that all objects experience the same density and
speed of impactors.  For instance, the mean semi-major axis of our
sample is $2.63 \pm 0.39(rms) \pm 0.04(err)$~AU compared to $2.806 \pm
0.300(rms) \pm 0.001(err)$~AU for the unbiased MB.  The difference is
because our sample of 95 asteroids came from SMASS that selected
bright (and therefore closer on average) asteroids for spectra
acquisition and because S-complex asteroids are primarily in the inner
MB.  The offset between the values is a significant fraction of the
width of the MB --- the spatial density of asteroids increases with
semi-major axis, impact speeds rise, and material strength decreases.
Additionally, S-complex asteroids have higher density than C-complex
asteroids which will affect their likelihood of disruption.  We hope
that future collisional evolution models will incorporate more detail
to allow a better comparison between space weathering and gardening
time estimates.

We also assumed that the space weathering rate is constant.  This
working assumption is reasonable even if the actual rate oscillates at
high frequency relative to our measured rate but could be problematic
if there is a secular trend or the rate oscillates slowly.  To explore
the possibility that the solar induced weathering rate changes with
time \citet{bib.rum09} proposes an expedition to investigate the lunar
regolith and plan to test their techniques on Hawaiian lavas.

This work has shown that age distributions from collisional evolution
and space weathering models are consistent.  Further refinement awaits
enhancements in the data sets and mechanisms for both models.  While
our result based on remote observation of asteroids suggests the
weathering time is long it is difficult to reconcile with some lab
results.  When this remaining discrepancy is resolved we will have
solved a four decade long search for the link between ordinary
chondrites and S-complex asteroids.

\section{Conclusion}\label{s.conclusion}

We used two independent methods to derive similar age distributions
for a sample of 95 S-complex asteroids from SMASS that also have $u,g,r,i,z$
filter magnitudes and absolute magnitudes from SDSS.  The first method
used absolute magnitudes to calculate diameters that are related to
age because large asteroids survive longer than small ones.  The
second method inverted a probabilistic relationship (the enhanced dual
$\tau$ model) between an asteroid's color and its age.  We developed
the enhanced dual $\tau$ model to avoid a color truncation problem
encountered when inverting the dual $\tau$ model of
\citet{bib.wil10}.  We then fit the color-age distribution from the
enhanced dual $\tau$ model to the size-age distribution and showed
that there is consistency between the two age determination
approaches.  This was not inevitable given the entirely independent
techniques and suggests we are converging on a self-consistent
understanding of both the collisional evolution and space weathering
models.

The most probable color for fresh S-complex asteroid surface is
$PC_1(0) = -0.05 \pm 0.01$ in agreement with the color of ordinary
chondrite meteorites of $PC_1(0) = 0.17 \pm 0.39$.  While we are
encouraged by the agreement between the two values we realize that it
is due to the large color range exhibited by the lab spectra of the
meteorites.  The wide variation in meteorite colors may be due to
differences in particle sizes or subtypes and the absolute difference
between our prediction and the color of meteorites may be resolved
with a better understanding of asteroid surface regolith.

According to our model the most probable color for S-complex asteroids
after infinite time in the absence of gardening is $PC_1(0) + \Delta
PC_1 = 1.29 \pm 0.04$.  The most probable ultimately attainable color
including the effect of gardening is $PC_1 = 0.91 \pm 0.07$.  Our
model indicates that most asteroids will not achieve this redness over
the age of the solar system.

The gardening time of $4400^{+700}_{-500}$ Myr derived here using the
enhanced dual $\tau$ model is over twice that found in
\citet{bib.wil10} that itself was about $7\times$ their calculated
resurfacing time.  It may be possible to reconcile the difference
using modifications to cratering phenomena proposed by
\citet{bib.wil10}.

Based on our small sample of 95 asteroids for which most span a narrow
size and age range of about $1-20$ km and $100 - 3000$ Myr
respectively we measured a space weathering time of $2050 \pm 80$ Myr.
This is much longer than some results based on particle bombardment
experiments that suggest weathering times of less than 1 Myr.  We are
unable to reconcile the discrepancy with the particle bombardment
experiments but it might indicate that protons or He ions are not the
primary cause of space weathering in the main belt.  We hope that in
the future larger data samples combined with improved collisional
evolution and space weathering models will solve the problem.

\section{Acknowledgments}

This work was supported under NSF grant AST04-07134.  We thank Mikael
Granvik and Bill Bottke for their helpful suggestions and insights.


\begin{thebibliography}

\bibitem[Abazajian \etal, 2009]{bib.aba09}
Abazajian, K. et al, 2009. The seventh data release of the Sloan
Digital Sky Survey. SDSS DR7
http://www.astro.washington.edu/ivezic/sdssmoc/sdssmoc.html

\bibitem[Adams and McCord, 1971]{bib.ada71}
Adams, J.B., McCord, T.B., 1971. Alteration of lunar optical
properties: age and composition effects. Science, 171, 567-571.

\bibitem[Asphaug \etal, 2002]{bib.asp02}
Asphaug, E., Ryan, E.V., Zuber, M.T., 2002. Asteroid interiors. In:
Bottke, W.F. Cellino, A., Paolicchi, P. Binzel, R.P., Editors,
2002. Asteroids III, Univ. of Arizona Press, Tucson, AZ, pp. 463-484.

\bibitem[Bowell, 2008]{bib.bow08}
Bowell, E., 2008.  Astorb. ftp://ftp.lowell.edu/pub/elgb/astorb.html

\bibitem[Binzel \etal, 2002]{bib.bin02}
Binzel, R.P., Dmitrij, F.L.,Martino, M.D., Whiteley, R.J., Hahn, G.J.,
2002. Physical properties of near-Earth objects. In: Bottke, W.F.,
Cellino, A., Paolicchi, P., Binzel, R.P. (Eds.), Asteroids III,
Univ. of Arizona Press, Tucson, pp. 255-271.

\bibitem[Binzel \etal, 2004]{bib.bin04}
Binzel, R.P., Rivkin, A.S., Stuart, J.S., Harris, A.W., Bus, S.J.,
Burbine, T.H., 2004. Observed spectral properties of near-Earth
objects: results for population distribution, source regions, and
space weathering processes. Icarus 170, 259-294.

\bibitem[Bottke \etal, 2005]{bib.bot05a} Bottke, W.F., Durda, D.D.,
Nesvorn\'{y}, D., Jedicke, R., Morbidelli, A., Vokrouhlick\'{y}, D.,
Levison, H., 2005. The fossilized size distribution of the main
asteroid belt. Icarus 175, 1, 111-140.

\bibitem[Bottke \etal, 2006]{bib.bot06}
Bottke, W., Chapman, C., 2006. Determining the Main Belt Size
Distribution Using Asteroid Crater Records and Crater Saturation
Models. In: Mackwell, S.,Stansbery, E. (Eds.), 37th Annual Lunar and
Planetary Science Conference, L.P.I.Tech.Report 37, 1349.

\bibitem[Brunetto \etal, 2006]{bib.bru06a}
Brunetto, R., Romano, F., Blanco, A., Fonti, S., Martino, M., Orofino,
V., Verrienti, C., 2006.  Space weathering of silicates simulated by
nanosecond pulse UV excimer laser.  Icarus 180, 546-554.

\bibitem[Brunetto \etal, 2006]{bib.bru06b} Brunetto, R., Vernazza, P.,
Marchi, S., Birlan, M., Fulchignoni, M., Orofino, V., Strazzulla, G.,
2006.  Modelling asteorid surfaces from observations and irradiation
experiments: the case of 832 Karin. Icarus 184, 327-337.

\bibitem[Bus and Binzel, 2002]{bib.bus02a}
Bus, S.J., Binzel, R.P., 2002. Phase II of the Small Main-Belt
Asteroid Spectroscopic Survey, The Observations. Icarus 158, 106-145.

\bibitem[Bus and Binzel, 2002]{bib.bus02b}
Bus, S.J., Binzel, R.P., 2002. Phase II of the Small Main-Belt
Asteroid Spectroscopic Survey, A Feature-Based Taxonomy. Icarus 158,
146-177.

\bibitem[Cellino \etal, 2001]{bib.cel01}
Cellino, A., Zappala, V., Doressoundiram, A., Di Martino, M., Bendjoy,
Ph., Dotto, E. Migliorini, F., 2001. The Puzzling Case of the
Nysa-Polana Family. Icarus 152, 225-237.

\bibitem[Chapman and Salisbury, 1973]{bib.cha73}
Chapman C.R., Salisbury, J.W., 1973. Comparison of meteorite and
asteroid spectral reflectivities. Icarus 19, 507-522.

\bibitem[Chapman \etal, 1975]{bib.cha75}
Chapman C.R., Morrison, D., Zellner, B., 1975.  Surface properties of
asteroids - A synthesis of polarimetry, radiometry, and
spectrophotometry. Icarus 25, 104-130.

\bibitem[Chapman, 2004]{bib.cha04}
Chapman, C.R., 2004. Space Weathering of Asteroid Surfaces,
Annu. Rev. Earth Planet. Sci., 32, 550-551.

\bibitem[Chapman, 2005]{bib.cha05}
Chapman, C., Merline, W., Thomas, P., Joseph, J., Cheng, A., Izenberg,
N., 2005. Impact History of Eros: Craters and Boulders, Icarus 155,
104-118.

\bibitem[Chapman \etal, 2007]{bib.cha07}
Chapman, C.R., Enke, B., Merline, W.J., Nesvorn\'{y}, D., Tamblyn,.P.,
Young, E.F., Olkin, C., 2007. Young Asteroid 832 Karin Shows no
rotational spectral variations. Icarus in press.

\bibitem[Clark \etal, 2002]{bib.cla02}
 Clark B. F., Hapke B., Pieters C., Britt D., 2002. Asteroid Space
Wearthing and Regolith Evolution. In: Bottke, W.F., Cellino, A.,
Paolicchi, P., Binzel, R.P. (Eds.), Asteroids III, Univ. of Arizona
Press, Tucson, pp. 585-599.

\bibitem[Cloutis, 1994]{bib.clo94}
Cloutis, E.A., 1994. Brown University Keck/NASA Relab Spectra Catalog,
http://lf314-rlds.geo.brown.edu/.

\bibitem[Dohnanyi, 1971]{bib.doh71}
Dohnanyi, 1971. Fragmentation and distribution of asteroids.  In:
Physical studies of minor planets, Gehrels, T., Editor, 1971. NASA
Special Publication 267, pp. 263-295.

\bibitem[Durda \etal, 1998]{bib.dur98}
Durada, D., Greenberg, R., Jedicke, R., 1998. Collisional models and
scaling laws: a new interpretation of the shape of the main-belt
asteroid size distribution. Icarus 135, 431-440.

\bibitem[Fevig and Fink, 2007]{bib.fev07}
Fevig, R., Fink, U., 2007. Spectral observations of 19 weathered and
23 fresh NEAs and their correlations with orbital parameters. Icarus
188, 175-188.

\bibitem[Fowler and Chillemi, 1992]{bib.fow92}
Fowler, J., Chillemi, J., 1992.  IRAS asteroid data processing. In:
Tedesco, E. (Ed.), The IRAS Minor Planet Survey.  Phillips Laboratory,
Nanscom Air Force Base, MA, 176-189. Tech. Report PL-TR-92-2049.

\bibitem[Gaffey \etal, 1993]{bib.gaf93}
Gaffey, M., Burbine, T., Piatek, J., Reed, K., Chaky, D., Bell, J.,
Brown, R., 1993.  Mineralogical variations within the S-type asteroid
class.  Icarus 106, 573-602.

\bibitem[Gladman \etal, 2000]{bib.gla00}
Gladman, B., Michel, P., Froeschle, C., 2000.  The Near-Earth Object
Population.  Icarus 146, 176-189.

\bibitem[Greenberg \etal, 1996]{bib.gre96}
Greenberg, R., Bottke, W., Nolan, M., Geissler, P., Petit, J., Durda,
D., Asphaug, E., Head, J., 1996. Collisional and dynamical history of
Ida. Icarus 120, 1, 106-118.

\bibitem[Hapke, 2001]{bib.hap01}
Hapke, B., 2000. How to turn OC's into S's: space weathering in the
asteroid belt. Lunar and Planetary Science, 31, 1087.

\bibitem[Hardorp, 1978]{bib.har78}
Hardorp, J., 1978. The Sun among the Stars. A\&A 63, 383-390.

\bibitem[Hinrichs and Lucey, 2002]{hin02}
Hinrichs, J.,Lucey, P., 2002. Temperature-Dependent Near-Infrared
Spectral Properties of Minerals, Meteorites, and Lunar Soil. Icarus
155, 169-180.

\bibitem[Hirayama, 1918]{bib.hir18}
Hirayama, K., 1918. Groups of asteroids probably of common origin. AJ
31, 185-188.

\bibitem[Hiroi \etal, 1996]{bib.hir96}
Hiroi, T., Vilas, F., Sunshine, J., 1996. Discovery and Analysis of
Minor Absorption Bands in S-Asteroid Visible Reflectance
Spectra. Icarus 119, 202-208.

\bibitem[Holsapple \etal, 2002]{bib.hol02}
Holsapple, K., Giblin, I., Housen, K., Nakamura, A., Ryan, E.,
2002. Asteroid impacts: laboratory experiments and scaling laws in
Asteroids III, Bottke, W., Cellino, A., Paolicchi, P. Binzel, R.,
editors, University of Arizona Press, Lunar and Planetary Institute.

\bibitem[Ivezi\'{c} \etal, 2000]{bib.ive00}
Ivezi\'{c}, \v{Z}., Goldston, J., Finlator, K., Knapp, G., Yanny, B.,
McKay, T., Amrose, S., Krisciunas, K., Willman, B., Anderson, S., and
32 others, 2000. Candidate RR Lyrae stars found in Sloan Digital Sky
Survey Commissioning Data. AJ, 120, 963-977.

\bibitem[Ivezi\'{c} \etal, 2002]{bib.ive02}
Ivezi\'{c}, \v{Z}., Juri\'{c}, M., Lupton, R. H., Tabachnik, S.,
Quinn, T., 2002. Asteroids Observed by The Sloan Digital Survey. In:
Survey and Other Telescope Technologies and Discoveries, Tyson, J.A.,
Wolff, S. (Eds.), Proc. SPIE, 4836, pp. 98-103.

\bibitem[JPL, 2006]{bib.jpl09}
JPL Small-Body Database, 2009. http://ssd.jpl.nasa.gov/sbdb.cgi.

\bibitem[Jedicke \etal, 1998]{bib.jed98}
Jedicke, R., Metcalfe, T., 1998. The orbital and absolute magnitude
distributions of main belt asteroids.  Icarus, 131, 245-260.

\bibitem[Jedicke \etal, 2004]{bib.jed04}
Jedicke, R., Nesvorn\'{y}, D., Whiteley, R.J., Ivezi\'{c}, \v{Z}.,
Juri\'{c}, M., 2004. An age-colour relationship for main-belt
S-complex asteroids.  Nature, 429, 275-277.

\bibitem[Krot \etal, 2003]{bib.kro05}
Krot, A.N., Keil, K., Goodrich, C.A., Scott, E.R.D., Weisberg, M.K.,
2005. Classification of Meteorites. In: Davis, A.M., Holland, H.D.,
Turekian, K.K., (Eds.), Meteorites, Comets and Planets, Vol. 1,
Treatise on Geochemistry, Elsevier, Oxford, pp. 86-116.

\bibitem[Loeffler \etal, 2009]{bib.loe09}
Loeffler, M., Dukes, C., Baragiola, R., 2009. Irradiation of olivine
by 4 keV He+: Simulation of space wethering by the solar wind.  JGR,
114, E03003.

\bibitem[Lugmair \etal, 1995]{bib.lug95}
Lugmair, G., Shukolyukov, A., MacIsaac, C., 1995. The Abundance of
$^{60}$Fe in the Early Solar System. In: Busso, M., Raiteri,
D.,Gallino, R., (Eds.), Nuclei in the Cosmos III, AIPCS 327, 591.

\bibitem[Marchi \etal, 2005]{bib.mar05}
Marchi, S., Brunetto, R., Magrin, S., Lazzarin, M., Gandolfi, D.,
2006. Space weathering of near-Earth and main belt silicate-rich
asteroids: observations and ion irradiation experiments. A\&A, 443,
769-775.

\bibitem[Marchi \etal, 2006]{bib.mar06}
Marchi, S., Paolicchi, P., Lazzarin, M., Magrin, S., 2006. A general
spectral slope-exposure relation for S-type main belt and near-earth
asteroids. AJ, 131, 1138-1141.

\bibitem[Marzari \etal, 1995]{bib.mar95}
Marzari, F.,Davis, D., Vanzani, V., 1995. Collisional evolution of
asteroid families. Icarus, 113, 168-187.

\bibitem[Masi \etal, 2007]{bib.mas07}
$http://people.roma2.infn.it/~masi/sdss_smass/sdsssmass.txt$

\bibitem[Moroz \etal, 1996]{bib.mor96}
Moroz, L., Fisenko, A., Semjonova, L., Pieters, C., Korotaeva, N.,
1996.  Optical Effects of Regolith Processes on S-Asteroids as
Simulated by Laser Shots on Ordinary Chondrite and Other Mafic
Materials. Icarus, 122, 366-382.

\bibitem[Nakamura \etal, 2001]{bib.nak01}
Nakamura, K., Sasaki, S., Hamabe, Y., Kurahashi, E., Hiroi, T., 2001.
Laboratory simulation of space weathering: A transmission electron
microscopic study - microstructures of the laser irradiated samples.
Lunar and Planetary Science Conference XXXII abstract \#1547.

\bibitem[Nesvorn\'{y} \etal, 2002]{bib.nes02}
Nesvorn\'{y}, D., Bottke, W.F., Dones, L., Levison, H.F., 2002. The recent
breakup of an asteroid in the main-belt region. Nature 417, 6890,
720-771.

\bibitem[Nesvorn\'{y} \etal, 2003]{bib.nes03}
Nesvorn\'{y}, D., Bottke, W.F., Levison, H.F., Dones, L. 2003. Recent
origin of the solar system dust bands. ApJ 591, 486-497.  720-771.

\bibitem[Nesvorn\'{y} \etal, 2005]{bib.nes05}
Nesvorn\'{y}, D., Jedicke, R., Whiteley, R.J., Ivezi\'{c}, \v{Z}.,
2005. Evidence for asteroid space weathering from the Sloan Digital
Sky Survey.  Icarus 173, 132-152.

\bibitem[Nesvorn\'{y} \etal, 2006a]{bib.nes06a}
Nesvorn\'{y}, D., Vokrouhlick\'{y}, D., Bottke, W.F., 2006. The Breakup of a
Main-Belt Asteroid 450 Thousand Years Ago. Science, 312, 1490.

\bibitem[Nesvorn\'{y} and Vokrouhlick\'{y}, 2006b]{bib.nes06b}
Nesvorn\'{y}, D., Vokrouhlick\'{y}, D., 2006. New Candidates for Recent
Asteroid Breakups. AJ, 132, 1950-1958.

\bibitem[Nesvorn\'{y} \etal, 2006c]{bib.nes06c}
Nesvorn\'{y}, D., Enke, B., Bottke, W.F., Durda, D., Asphaug, E.,
Richardson, D., 2006. Karin cluster formation by asteroid
impact. Icarus, 183, 296-311.

\bibitem[Paolicchi \etal, 2007]{bib.pao07}
Paolicchi, P., Marchi, S., Nesvorny, D., Magrin, S., Lazzarin, M.,
2007.  Towards a general model of space weathering of S-complex
asteroids and ordinary chondrites. A\&A 464, 1139-1146.

\bibitem[Pieters \etal, 1994]{bib.pie94}
Pieters, C.M., Staid, M.I., Fischer, E.M., Tompkins, S., He, G., 1994.
A Sharper View of Impact Craters from Clementine Data. Science 266,
5192, 1844-1848.

\bibitem[Pieters \etal, 2000]{bib.pie00}
Pieters, C.M., Taylor, L.A., Noble, S.K., Lindsay, P.K., Hapke, B.,
Morris, R.V., Allen, C.C., McKay, D.S., Wentworth, S., 2000. Space
weathering on airless bodies: Resolving a mystery with lunar samples.
Meteorit. Planet. Sci. 35, 1101-1107.

\bibitem[Pravec \etal, 2009]{bib.pra09b}
Pravec, P., Vokrouhlicky, D., 2009. Significance analysis of asteroid
pairs.  Icarus 204, 580-588.

\bibitem[Rumpf \etal, 2009]{bib.rum09}
Rumpf, E., Fagents, S., Crawford, I., Joy, K., 2009. The preservation
of ancient solar wind particles buried beneath lunar basalt flows as
determined through heat transfer modeling.  AGU Fall Meeting
Abstracts, C1290+.

\bibitem[Sasaki \etal, 2001]{bib.sas01}
Sasaki, S., Nakamura, K., Hamabe, Y., Kurahashi, E., Hiroi, T., 2001.
Production of Iron Nanoparticles by Laser Irradiation in a Simulation
of Lunar-like Space Weathering. Nature, 410, 555-557.

\bibitem[Scott \etal, 2006]{bib.sco06}
Scott, E., 2006.  Meteoritical and dynamical constraints on the growth
mechanisms and formation times of asteroids and Jupiter, Icarus 185,
72-82.

\bibitem[Scott \etal, 2007]{bib.sco07}
Scott, E., 2007.  Chondrites and the Protoplanetary Disk, AREPS 35,
577-620.

\bibitem[Scott \etal, 2009]{bib.sco09}
Scott, E., Bogard, D., Bottke, W., Taylor, G., Greenwood, R., Franchi,
I., Keil, K., Moskovitz, N., Nesvorn{\'y}, D., 2009.  Impact Histories
of Vesta and Vestoids Inferred from Howardites, Eucrites, and
Diogenites. In: Lunar and Planetary Institute Science Conference
Abstracts, L.P.I.Tech.Report 40, 2295.

\bibitem[Sheinis \etal, 2002]{bib.she02}
Sheinis, A.I., Bolte, M., Epps, H.W., Kibrick, R.I., Miller, J.S.,
Radovan, M.V., Bigelow, B.C., Sutin, B.M., 2002. ESI, A New Keck
Observatory Echellette Spectrograph and Imager. PASP, 114, 851-865.

\bibitem[Srinivasan \etal, 2002]{bib.sri99}
Srinivasan, G., Goswami, J., Bhandari, N., , 1999. 26Al in Eucrite
Piplia Kalan: Plausible Heat Source and Formation Chronology. Science,
284, 1348.

\bibitem[Stoughten, 2002]{bib.sto02}
Stoughten, C., 191 colleagues, 2002. Sloan Digital Sky Survey: early
data release. AJ 123, 485-548.

\bibitem[Strazulla \etal, 2005]{bib.str05} Strazzulla, G., Dotto, E.,
Binzel, R., Brunetto, R., Barucci, M.A., Blanco, A., Orofino, V.,
2005.  Spectral alteration of the meteorite Epinal (H5) induced by
heavy ion irradiation: A simulation of space weathering effects on
near-Earth asteroids, Icarus 174, 31-35.

\bibitem[Tholen, 1984]{bib.tho84}
Tholen, D.J., 1984. Ph.D. Dissertation. Univ. of Arizona, p. 95.

\bibitem[Vernazza \etal, 2009]{bib.ver09}
Vernazza, P., Binzel, R., Birlan, Fulchignoni,M., Rossi, A., 2009.
Solar wind as the origin of rapid reddening of asteroid
surfaces. Nature 458, 993-995.

\bibitem[Veverka \etal, 1996]{bib.vev96}
Veverka, P., Helfenstein, P., Lee, P., Thomas, P., McEwen, A., Belton,
M., Klaasen, K., Johnson, T., Granahan, J., Fanale, F., Geissler, P.,
Head, J., 1996.  Ida and Dactyl: spectral reflectance and color
variations.  Icarus 120, 66-76.

\bibitem[Vokrouhlick\'{y} \etal, 2006a]{bib.vok06a}
Vokrouhlick\'{y}, D., Broz, M., Bottke, W.F., Nesvorn\'{y}, D., Morbidelli,
A., 2006. Yarkovsky/YORP chronology of asteroid families. Icarus 182,
118-142.

\bibitem[Vokrouhlick\'{y} \etal, 2006b]{bib.vok06b}
Vokrouhlick\'{y}, D., Broz, M., Morbidelli, A., Bottke, W.,
Nesvorn\'{y}, D., Lazzaro, D., Rivkin, A., 2006. Yarkovsky footprints
in the Eos family. Icarus 182, 92-117.

\bibitem[Vokrouhlick\'{y} and Nesvorn\'{y}, 2008]{bib.vok08}
Vokrouhlick\'{y}, D., Nesvorn\'{y}, D., 2008.  Pairs of asteroids
probably of a common origin. AJ, 136, 380-290.

\bibitem[Willman \etal, 2008]{bib.wil08} Willman, M., Jedicke, R.,
Nesvorny, D., Moskovitz, N., Ivezi\'{c}, \v{Z}., Fevig, R., 2008.
Redetermination of the space weathering rate using spectra of Iannini
asteroid family members. Icarus 195, 663-673.

\bibitem[Willman \etal, 2010]{bib.wil10}
Willman, M., Jedicke, R., Moskovitz, N., Nesvorn\'{y}, D.,
Vokrouhlick\'{y}, D., Moth\'{e}-Diniz, T., 2010.  Using the youngest
asteroid clusters to constrain the Space Weathering rate on S-complex
asteroids. Icarus 208, 758-772.

\end{thebibliography}
\end{document}